\documentclass[a4paper]{jpconf}
\usepackage{graphicx}
\begin{document}
\title{Perspectives for Top Quark Physics at the (I)LC}

\author{Frank Simon}

\address{Max-Planck-Institut f\"ur Physik, F\"ohringer Ring 6, 80805 M\"unchen, Germany}

\ead{fsimon@mpp.mpg.de}

\begin{abstract}
Linear $e^+e^-$ colliders provide a rich set of opportunities for  top  quark physics, crucial for the understanding of electroweak symmetry breaking and for the search for physics beyond the Standard Model. A $t\bar{t}$ threshold scan in $e^+e^-$ annihilation enables a precise measurement in theoretically well-defined mass schemes with small experimental and theoretical systematic uncertainties. Above the production threshold, the efficient identification of top pair events combined with polarized beams provides the potential to extract the form factors for the top quark couplings with high precision and in a model-independent way, resulting in excellent sensitivity to physics beyond the Standard Model. This contribution provides an overview of top physics at linear colliders based on results from full-simulation studies of top quark pair production in the detectors proposed for ILC and CLIC. In addition, the influence of the luminosity spectrum of a 100 km circular $e^+e^-$ collider (FCCee) compared to linear colliders on a top threshold scan is briefly discussed.
\end{abstract}

\section{Introduction}

As the heaviest particle in the Standard Model, the top quark has a special role. Due to its high mass, it has the the strongest coupling of all known particles to the Higgs field. It also takes a central role in many models of New Physics, and thus provides a high sensitivity for phenomena beyond the Standard Model.

To date, the top quark is the only quark that has been studied  exclusively at hadron colliders. The clean experimental environment in $e^+e^-$ collisions enables the study of all decay modes of the top quark with high resolution and very low background levels. The measurements at lepton colliders also profit from the high precision of theoretical calculations, which result in small overall systematic uncertainties. 

At $e^+e^-$ colliders, there are two different main programs for top physics. The first is the study of the threshold for top quark pair production, which provides access to the detailed properties of the top quark. The second is the use of the top quark as a tool for the search for physics beyond the Standard Model, for example by precisely measuring its coupling to the electroweak interactions. In the first case, collision energies at several different values around 350 GeV are necessary, while the second program requires energies substantially in excess of the threshold for top pair production, of the order of 500 GeV or higher. In particular, the energies above threshold are uniquely available at linear $e^+e^-$ colliders.

Two such high-energy $e^+e^-$ colliders are currently being developed in international collaboration, the International Linear Collider (ILC)  \cite{Behnke:2013xla} and the Compact Linear Collider (CLIC) \cite{Lebrun:2012hj}.  They are based on different acceleration technologies, resulting in a different energy reach for the full projects. ILC is based on superconducting RF structures, and is planned as a 500 GeV collider with operation at different energies from 250 GeV to 500 GeV, including the region around the $t\bar{t}$ threshold, and the possibility for upgrades to one TeV. CLIC uses a normal-conducting two-beam acceleration scheme, and is foreseen to be constructed in several stages. It has an ultimate energy of 3 TeV and two lower energy stages to maximise the physics potential, with the first stage covering the $t\bar{t}$ threshold. For ILC, the technical design report has been completed, while for CLIC a conceptual design report was delivered, with a technical design phase still ongoing until 2018.

In the following, the top physics program at these future colliders is illustrated based on two examples that have been studied with detailed simulations with realistic detector models, including physics and machine-related backgrounds. For the $t\bar{t}$ threshold scan, studies have been performed for ILC and CLIC, while the investigation of the physics potential for measurements of the electroweak couplings of the top quark have been performed for ILC at \mbox{500 GeV}. 

\section {A top threshold scan at ILC and CLIC}

The cross section of $t\bar{t}$ production close to the threshold strongly depends on the top quark mass. In addition, it receives contributions from the top quark width, from the strong coupling and from the top Yukawa coupling. The top width influences the shape of the would-be bound state of the $t\bar{t}$ pair. The strong coupling and the top Yukawa coupling, which both influence the interaction of the two top quarks, primarily affect the overall magnitude of the cross section. Beyond those effects connected to the $t\bar{t}$ system, the cross section also receives corrections due to initial state radiation (ISR) and due to the luminosity spectrum of the collider. The pure $e^+e^- \to t\bar{t}$ cross section can be calculated with high precision, resulting in clean theoretical predictions for the observables based on theoretically well-defined parameters, such as the 1S mass of the top quark. 

\begin{figure}[ht]
\centering
\includegraphics[width=0.48\textwidth]{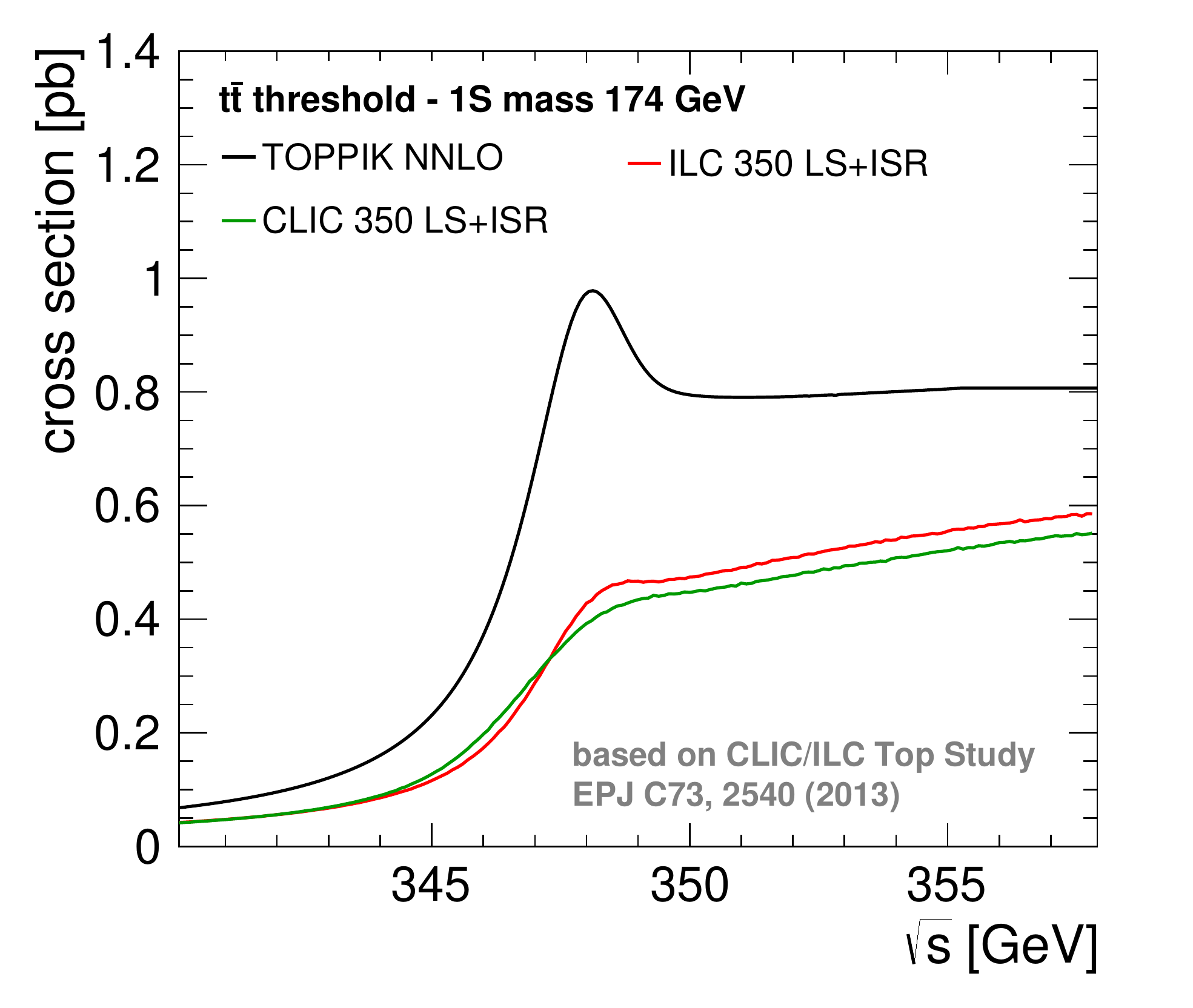}
\hfill
\includegraphics[width=0.48\textwidth]{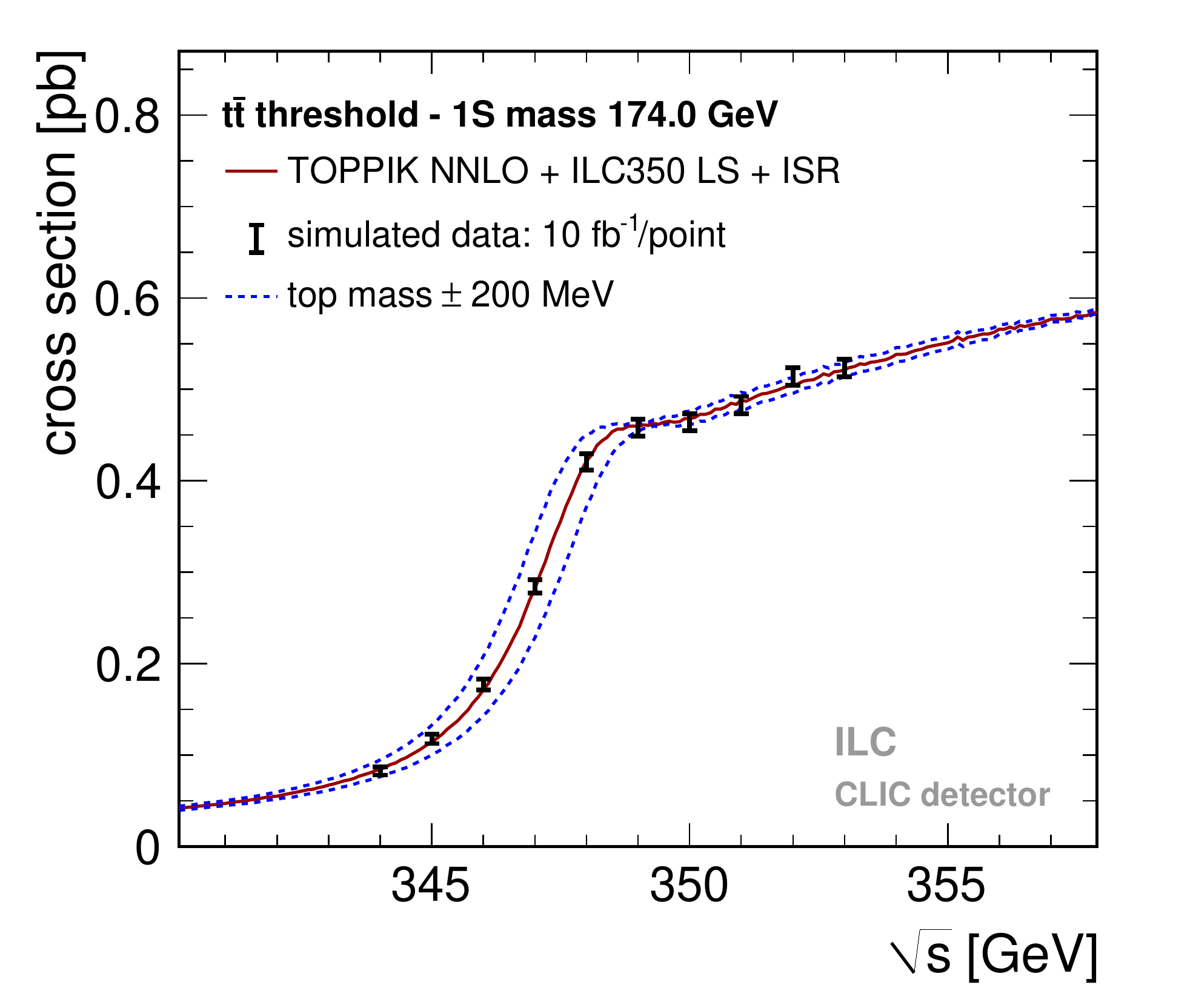}
\caption{The total $t\bar{t}$ cross section in the threshold region based on TOPPIK NNLO calculations \cite{Hoang:1999zc, Hoang:1998xf} including ISR and luminosity spectrum effects for ILC and CLIC ({\it{left}}) and a simulated $t\bar{t}$ threshold scan at ILC with 10 points spaced by 1 GeV each, assuming an integrated luminosity of 10 fb$^{-1}$ per point with unpolarised beams ({\it{right}}). For illustration purposes, the effect of a shift in the 1S top quark mass by $\pm$200 MeV is shown in addition to the simulated data points. Figure taken from \cite{Seidel:2013sqa} .}\label{fig:LS_Threshold}
\end{figure}

Figure \ref{fig:LS_Threshold} {\it{left}} illustrates the effect of the different luminosity spectra of the two machines together with initial state radiation on the pure $t\bar{t}$ production cross section calculated with NNLO QCD \cite{Hoang:1999zc, Hoang:1998xf}. The luminosity spectrum and ISR result in an overall reduction of the effective cross section since they shift a fraction of the luminosity below the threshold energy, and lead to a broadening of the threshold turn-on due to the low-energy tail and due to the width of the main luminosity peak. Since the beam energy spread is larger at CLIC than at ILC, the smearing is slightly more pronounced at CLIC. 

With full detector simulations of the CLIC\_ILD concept \cite{Linssen:2012hp} the reconstruction efficiencies for $t\bar{t}$ events and the rejection efficiency for Standard Model background was determined in the threshold region. These efficiencies are used to simulate threshold scans at ILC and CLIC, as illustrated in Figure \ref{fig:LS_Threshold} {\it{right}}. From these scans, the top mass is determined via a template fit of the measured cross section. With a 10-point scan with a total integrated luminosity of \mbox{100 fb$^{-1}$} assuming unpolarised beams, the top quark mass can be determined with a statistical uncertainty of 18 MeV in the case of ILC and 21 MeV in the case of CLIC \cite{Seidel:2013sqa}. The current precision of $\alpha_s$ of 0.0007 leads to a systematic uncertainty of equal magnitude (18 MeV for ILC, 20 MeV for CLIC), which is expected to improve in the future with a more precise determination of the strong coupling constant. 

In addition, there are experimental systematic uncertainties from several sources, such as the beam energy and the reconstruction efficiency and background contamination, which are expected to have a total size below 50 MeV. On top of that, there are theoretical systematics due to the precision of the calculation of the total $t\bar{t}$ cross section. These depend on the details of the calculations, and are still being evaluated. With the simplified assumption of a 3\% overall normalisation uncertainty, the resulting systematic uncertainty of the mass is around 50 MeV, making theoretical uncertainties potentially the leading source of systematics. 

A potentially important source of experimental systematics is the knowledge of the luminosity spectrum. The impact of this has been studied for CLIC by reconstructing the luminosity spectrum from simulated measurements of large-angle Bhabha scattering \cite{Poss:2013oea}. In a preliminary study, the uncertainty resulting from the precision of the reconstructed luminosity spectrum has been found to be of the order of 6 MeV, demonstrating that this is not a limiting factor for the overall precision of the top quark mass measurement. Since the luminosity spectrum at ILC is less complicated than the one at CLIC, even smaller uncertainties are expected for the ILC case. 

Overall, a $t\bar{t}$ threshold scan at linear colliders is expected to provide the top quark mass in a theoretically well-defined mass scheme with sub-100 MeV total uncertainty, which would provide a knowledge of the $\overline{MS}$ mass of the top quark at the 100 MeV level or better. Given the dominance of systematics and the uncertainties involved in the conversion from the 1S to the  $\overline{MS}$ mass, the small differences in statistical uncertainties between the different linear collider options are insignificant. 

\subsection{A top threshold scan at FCCee}

With the recent start of a design study for a 80 km to a 100 km circular collider at CERN, which could ultimately provide proton-proton collisions at center-of-mass energies of up to 100 TeV, the possibilities for an $e^+e^-$ collider, FCCee \cite{Gomez-Ceballos:2013zzn}, in the same tunnel as a precursor to the hadron collider, are being explored. This collider could reach up to the $t\bar{t}$ threshold, and thus would be capable of measuring the top quark mass in a threshold scan.

\begin{figure}[ht]
\centering
\includegraphics[width=0.48\textwidth]{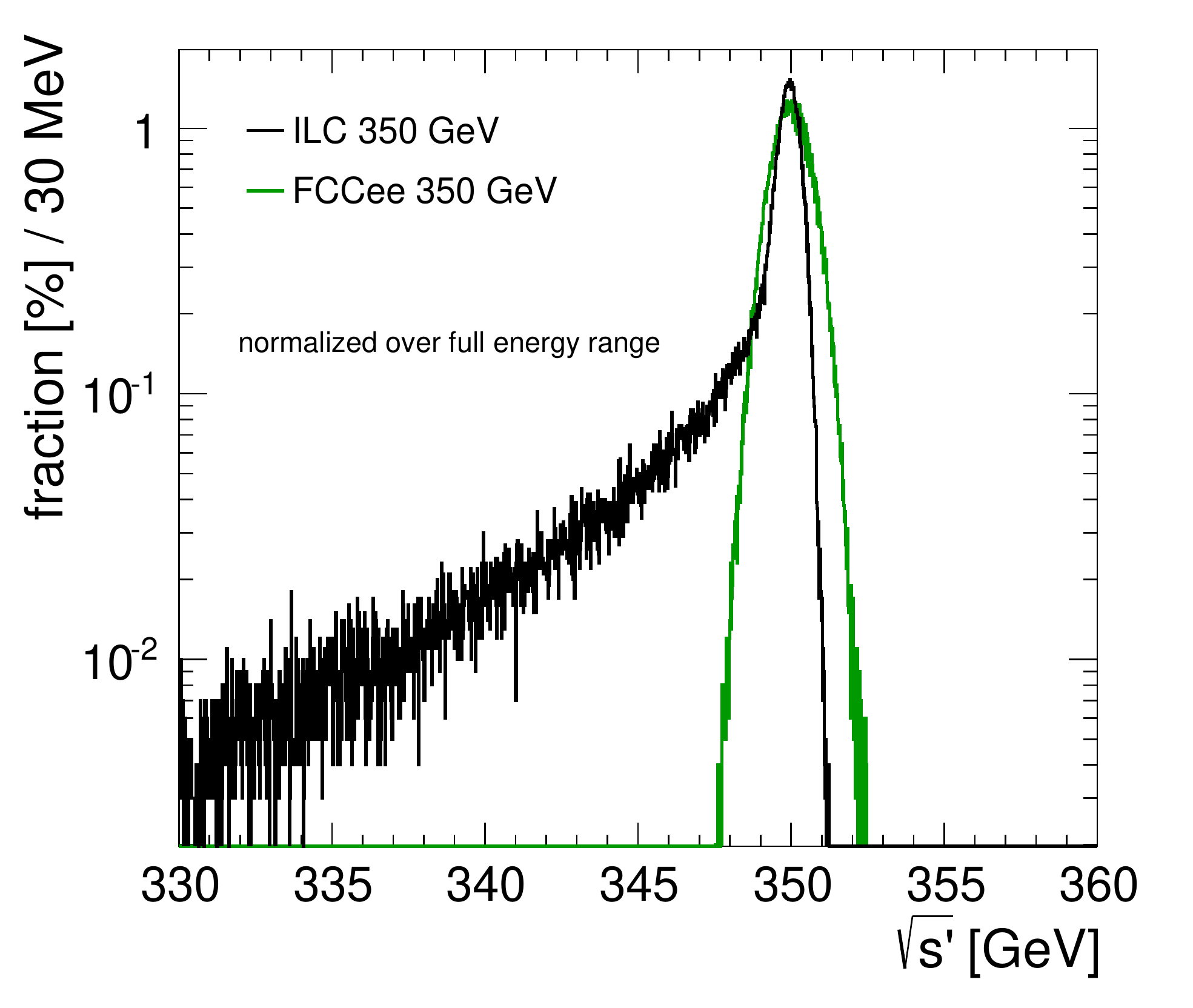}
\hfill
\includegraphics[width=0.48\textwidth]{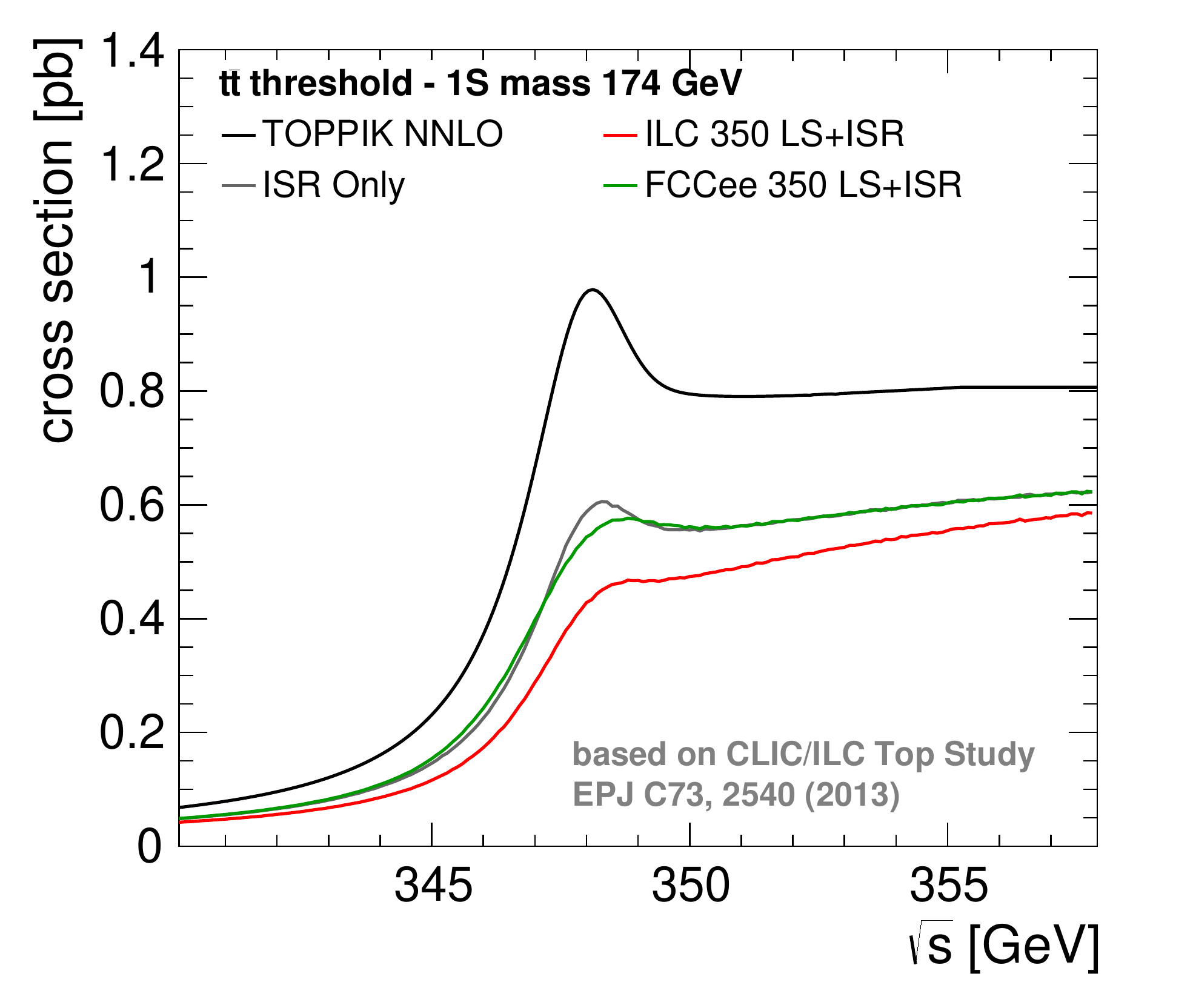}
\caption{The luminosity spectrum for ILC and FCCee at 350 GeV ({\it left}) and the resulting total $t\bar{t}$ cross section in the threshold region based on TOPPIK NNLO calculations \cite{Hoang:1999zc, Hoang:1998xf} including ISR and luminosity spectrum effects. Unpolarised beams are assumed. The capability for polarised beams at ILC can be used to increase the cross section or to reduce backgrounds.}\label{fig:FCC_Threshold}
\end{figure}

Figure \ref{fig:FCC_Threshold} {\it{left}} shows the luminosity spectrum of FCCee compared to ILC, assuming a gaussian energy spread with a sigma of 0.19\%, based on the RMS beam energy spread of the current FCCee design for 350 GeV including beamstrahlung \cite{Zimmermann:2014qxa}. The effect of this luminosity spectrum on the $t\bar{t}$ threshold is illustrated in Figure  \ref{fig:FCC_Threshold} {\it{right}} in comparison to the situation at ILC. The absence of the low-energy beamstrahlungs tail at FCCee leads to a larger effective cross-section compared to ILC, but in contrast to initial expectations \cite{Gomez-Ceballos:2013zzn} the corrections due to the luminosity spectrum result in a substantial smearing of the peak of the cross-section, which is more pronounced than at ILC due to the larger beam energy spread. Assuming the same detector performance as for linear colliders with identical top quark identification and background rejection, a 10 point scan with an integrated luminosity of 100 fb$^{-1}$ results in a 16 MeV statistical uncertainty of the top mass, a reduction by 10\% compared to ILC due to the somewhat higher event statistics. For all colliders, these uncertainties scale with $\sqrt{\int \cal{L}}$, and can be further reduced by increased running time at the threshold. Given the dominance of systematic uncertainties, in particular theory (including $\alpha_s$) uncertainties when converting to the $\overline{MS}$ mass of the top quark, the slight difference in statistical uncertainties of circular and linear colliders, as well as potentially substantially increased integrated luminosities will likely not lead to differences in the achievable total precision.

\section{Electroweak couplings}

The capability for polarised beams at linear colliders provides excellent conditions to probe the electroweak couplings of the top quark in $t\bar{t}$ production above threshold. These couplings are precisely determined in the Standard Model, but may receive substantial modifications in scenarios with physics beyond the Standard Model, such as extra dimensions and Higgs compositeness. The measurement of the total production cross section, the forward-backward asymmetry and the helicity angle, each for two different polarisation configurations, provides sufficient information to fully constrain the top quark couplings with high precision. Since the asymmetry and angle measurements require the identification of the top quark charge and rely on the correct association of $W$ bosons and $b$ jets to top candidates, these measurements profit from higher energy which provides a clean separation of the two top quarks in the $t\bar{t}$ system. 

\begin{figure}[ht]
\centering
\includegraphics[width=0.55\textwidth]{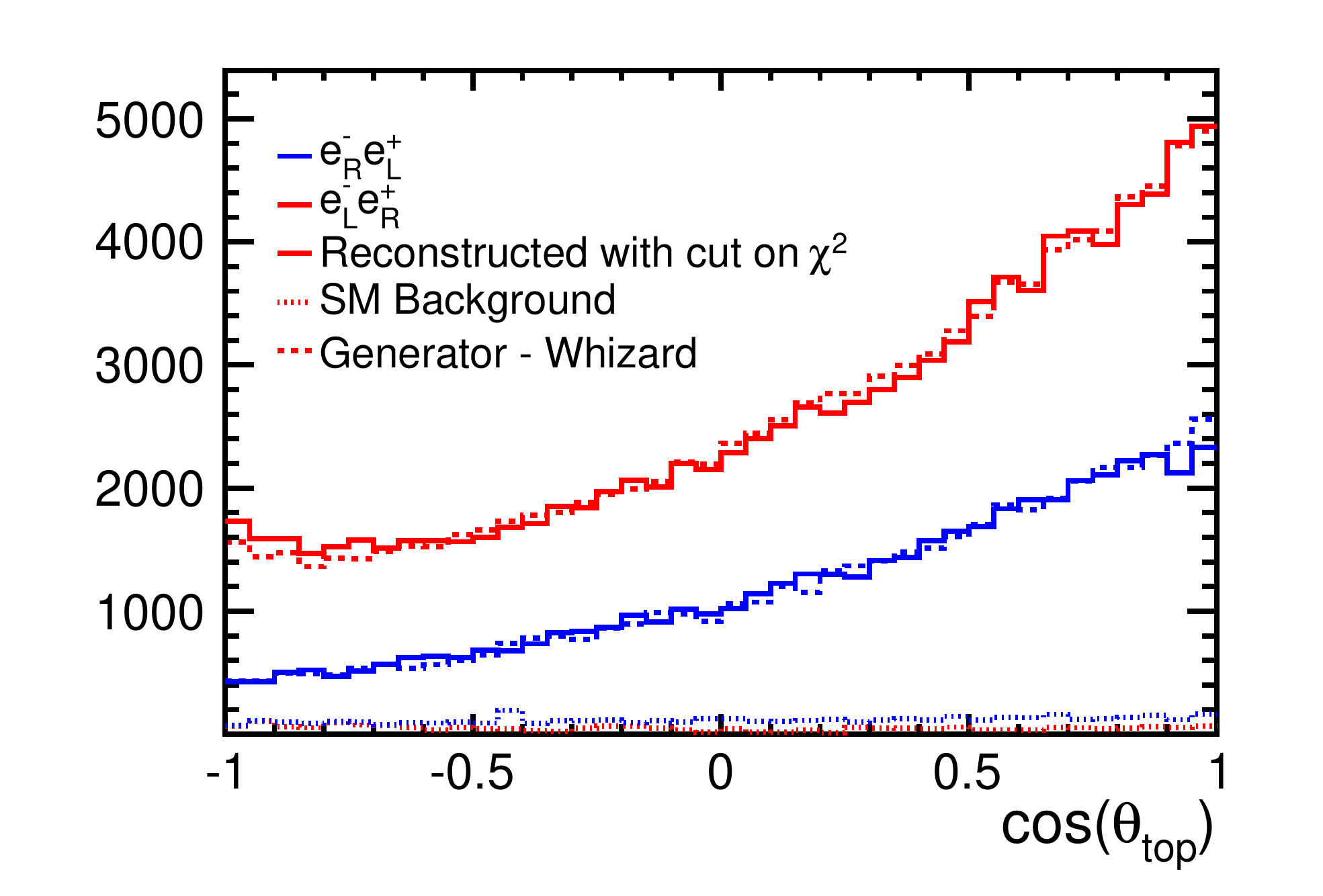}
\hfill
\includegraphics[width=0.42\textwidth]{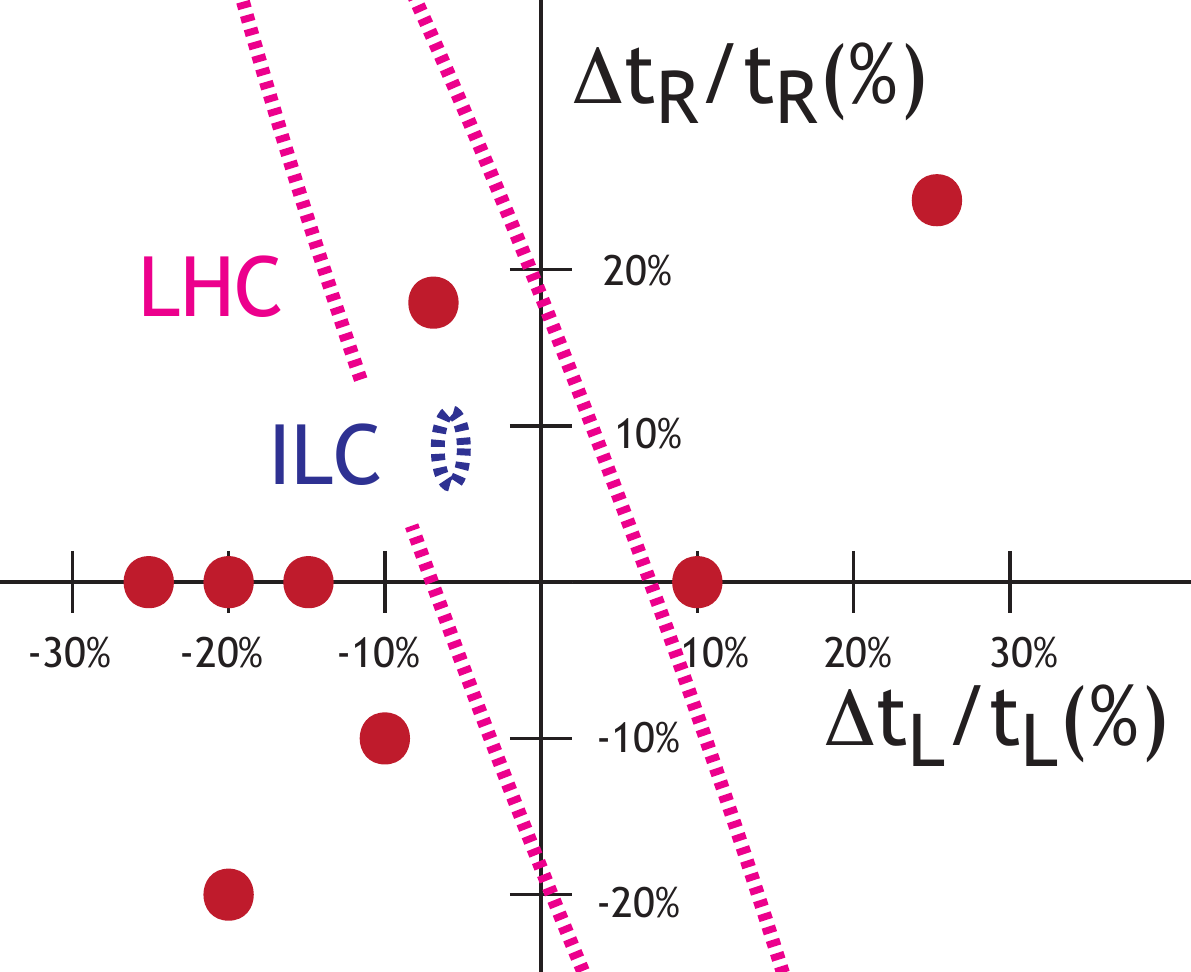}
\caption{The forward-backward asymmetry of $t\bar{t}$ production at 500 GeV in a full simulation study for two different polarisation configurations of the beams ($\pm$ 80\% $e^-$, $\mp$ 30\% $e^+$ polaristion) ({\it left}). Figure taken from \cite{Amjad:2013tlv}. The precision achievable at ILC at 500 GeV with an integrated luminosity of 500 fb$^{-1}$ for the left- and right-handed couplings of the top quark, compared to the expected precision achievable at the HL-LHC, together with the deviations from the Standard Model values in a variety of models with composite Higgs bosons \cite{Richard:2014upa, Agashe:2013hma} ({\it right}).}\label{fig:Asymmetry}
\end{figure}

Figure \ref{fig:Asymmetry} {\it left} shows the  forward-backward asymmetry in a full simulation study with the ILD detector \cite{Behnke:2013lya} for an integrated luminosity of  500 fb$^{-1}$ at an energy of 500 GeV at ILC. The total integrated luminosity is equally split between two polarisation configurations of $\pm$80\%, $\mp$30\% for electrons and positrons, respectively \cite{Amjad:2013tlv}.  From this measurement, the forward-backward asymmetry is extracted with a $\sim$2\% uncertainty including statistical and systematic contributions. Similarly, the helicity angle distribution can be extracted with a precision of 4\%, and the total cross section with a 0.5\% uncertainty. From these results, the left- and right-handed couplings can be extracted with a 0.7\% and 1.8\% precision, respectively \cite{Amjad:2013tlv, Richard:2014upa}. This precision is illustrated in Figure  \ref{fig:Asymmetry} {\it right} together with the predicted deviations from the Standard Model for several scenarios of New Physics with composite Higgs bosons and with the precision expected from the HL-LHC \cite{Richard:2014upa, Agashe:2013hma}. This clearly illustrates the immense power of a polarised high-energy electron-positron collider not only to discover possible new phenomena in the top sector, but also to precisely pin down the underlying mechanism if deviations from the Standard Model are observed. 

\section{Summary}

The future linear electron-positron colliders ILC and CLIC provide excellent opportunities for a precise study of the top quark sector. With polarised beams, the possibility for a scan of the $t\bar{t}$ production threshold and with measurements of the electroweak couplings of the top quark at energies substantially above the threshold they will provide high precision measurements of top quark properties and significant sensitivity for various physics scenarios beyond the Standard Model. A linear collider will  determine the top-quark mass in the theoretically well-defined $\overline{MS}$ scheme with a total precision of 100 MeV or better, and is capable of a percent-level measurement of the top electroweak couplings, which provides sensitivity to new physics scales extending substantially beyond the direct reach of present colliders.

\end{document}